\documentstyle[twocolumn,prl,aps]{revtex}
\large

\begin{document}
\draft
\twocolumn[
\hsize\textwidth\columnwidth\hsize\csname @twocolumnfalse\endcsname
\title{
	Phonon-like and single particle dynamics in liquid lithium 
	}
\author{
	T.~Scopigno$^{1}$,
	U.~Balucani$^{2}$,
	A.~Cunsolo$^{3}$,
	C.~Masciovecchio$^{3}$,
	G.~Ruocco$^{4}$,
	F.~Sette$^{3}$,	
	R.~Verbeni$^{3}$	
	}
\address{
	 $^{1}$ Universit\'a di Trento and Istituto Nazionale 
	 di Fisica della Materia, I-38100, Trento, Italy.\\
	 $^{2}$ Istituto di Elettronica Quantistica, 
         Consiglio Nazionale delle Ricerche, I-50127, Firenze, Italy.\\
	 $^{3}$ European Synchrotron Radiation Facility. 
	 B.P. 220 F-38043 Grenoble, Cedex France. \\
	 $^{4}$ Universit\'a di L'Aquila and Istituto Nazionale 
	 di Fisica della Materia, I-67100, L'Aquila, Italy.\\
	}
\date{\today}
\maketitle

\begin{abstract}
The dynamic structure factor, $S(Q,E)$, of liquid lithium ($T$=475 K)
has been
determined  by inelastic x-ray scattering (IXS) in the momentum transfer 
region ($Q$= 1.4$\div$110 nm$^{-1}$). These data allow to observe how, in 
a simple liquid, a phonon-like collective mode evolves towards the single 
particle dynamics. As a function of $Q$, one finds: {\it i)}at low $Q$'s, a 
sound mode with a positive dispersion of the sound velocity, {\it ii)} at 
intermediate $Q$'s, excitations whose energy oscillates similarly to phonons
in the crystal Brillouin zones, and {\it iii)} at high $Q$'s, the $S(Q,E)$
approaches a Gaussian shape, indicating that the single particle dynamics 
has been reached.

\end{abstract}
] 

The spectrum of density fluctuations, $S(Q,E)$, of a liquid shows a very
rich phenomenology with features strongly dependent on the considered
momentum, $Q$, and energy, $E$, regions \cite{BY,baluc}. The shape of the 
$S(Q,E)$ is well established in the small and high $Q$ limits. In the
hydrodynamics limit, i.~e. at $Q/Q_M<<1$ ($Q_M$ is the position of the first
maximum of the static structure factor $S(Q)$), the spectrum shows three
peaks - the Brillouin triplet. They are respectively the Stokes and
AntiStokes propagating compressional modes -dispersing linearly with the
adiabatic sound velocity $v_s$, and the thermal diffusion mode centered at
zero energy transfer. In the opposite limit, i.~e. at $Q/Q_M>>1$, one
reaches the so called impulse approximation, where the excited particles
acquires a kinetic energy much higher than any inter-particle potential
energy. Therefore the target particle behaves as a free particle, and the 
$S(Q,E)$ lineshape reflects the particles' initial state momentum
distribution. Considering the Boltzmann distribution, the $S(Q,E)$ reduces
to a Gaussian centered at the recoil energy $\hbar ^2Q^2/2M$, and with
variance $\hbar^2 K_BTQ^2/M$. Here, $M$ is the particel mass. The evolution 
between these two limit cases is affected by a realm of dynamical 
processes as the interaction of sound waves with other degrees of freedom 
(translational diffusion, rotations and internal modes for molecules), and 
the interaction between different collective modes, responsible, for example, 
for the slowing down of the diffusional dynamics in supercooled liquids. 
Moreover, when $Q$ is comparable to $Q_M$, important modifications of the 
Brillouin triplet occur also as a consequence of structural effects.
In fact, the sound waves in the liquid cannot be longer described as
density fluctuation
of a continuum medium, and the local structure with its intrinsic disorder
becomes relevant. Compared to the crystalline case, the absence of long range
order will prevent to a certain extent, in the liquid, the replication of
the sound dispersion relation in high order Brillouin Zones (BZ).

This rich phenomenology has motivated, since long time, the experimental
study of the dynamics of liquid systems. Using ultrasound absorption methods
and Brillouin light scattering techniques, the sound waves, and their
interactions with the relaxation processes active in the liquid, have been
studied in great detail up to $Q/Q_M\approx 10^{-3}$. Similarly, Inelastic
Neutron Scattering (INS) has been used at very large $Q$ transfers to
determine the $S(Q,E)$ in the impulse approximation, and in $Q$ regions
close to $Q_M$ to determine the $S(Q,E)$ lineshape when the two limit 
description are expected to fail \cite{jongth,jong}. The kinematic
limitations on the accessible $Q-E$ region of existing INS instruments \cite
{loves} did not allow to study with continuity and under comparable
experimental conditions the liquid dynamics in the $Q$ range spanning from
the collective to the single particle behaviors. Moreover the neutron 
scattering cross-section accounts for two different contributions:
besides the coherent cross section -probing the collective dynamics- 
there is an incoherent contribution which, at each wavevector, reflects 
the single particle motion and therefore hides the crossover between the
previously quoted regimes. The devolepment of Inelastic X-ray Scattering 
(IXS) \cite{burk} has allowed, recently, to sensibly extend the accesible 
$Q-E$ region in disordered materials and to avoid the incoherent 
contribution.

In this work we report the determination by IXS of the dynamic structure
factor of liquid lithium in the 1.4$\div $110 nm$^{-1}$ $Q$ range,
corresponding to $Q/Q_M\approx 5\cdot 10^{-2}-5$. Lithium has been chosen
because, among the simple monoatomic liquids, is the one that is best suited
to IXS. Indeed, its low mass gives recoil energies observable in the
considered $Q$-range, and its low atomic number and large sound velocity 
($\approx $5000 m/s) give optimal signal with the available energy resolution,
compensating for the large form factor decrease at high $Q$ values \cite
{oshe}. The $S(Q,E)$ spectra, reported to their absolute scale exploiting
the zeroth and first moments sum rules, show the transition from a triplet
to a Gaussian. The maxima, $\Omega (Q)$, of the longitudinal current spectra
($E^2/Q^2S(Q,E)$) show an almost linear dispersion relation at low $Q$ -
typical of a sound wave - and a completely different dependence in the high 
$Q$ limit, where it approaches the parabolic dispersion of the free particle.
These two regions are joined by oscillations of $\Omega (Q)$, which are in
phase with the structural correlations, as observed in the $S(Q)$. Finally,
in the low $Q$ region these data confirm the existence of positive
dispersion of the sound velocity.

The experiment has been carried out at the very high energy resolution IXS
beamline ID16, at the European Synchrotron Radiation Facility. The incident
x-ray beam is obtained by a back-scattering monochromator operating at the
Si(h~h~h) (h=7, 9, 11) reflections \cite{noiV}. The scattered photons are
collected by spherical silicon crystal analyzers, operating at the same
Si(h~h~h) reflection \cite{noiM}. The total energy resolution $-$ obtained
from the measurement of $S(Q_M,E)$ in a Plexiglas sample which is dominated
by elastic scattering $-$ is 8.5~meV full-width-half-maximum for h=7, 3~meV
for h=9 and 1.5~meV for h=11. The momentum transfer,
$Q=2k_h sin(\theta_s/2)$, with $k_h$ the wavevector of
the incident photon and $\theta_s$ the
scattering angle, is selected either between 1.4 and 25~nm$^{-1}$ by
rotating a 7~m long analyser arm in the horizontal scattering plane (data
taken at h=9 and 11), or between 24 and 110~nm$^{-1}$ by rotating a 3~m long
analyser arm in the vertical scattering plane (h=7). The total $Q$
resolution has been set to 0.4~nm$^{-1}$ at h=9 and 11, and 1~nm$^{-1}$ at
h=7. On the horizontal arm, five independent analyzers were used to collect
simultaneously five different $Q$ values, determined by the constant angular
offset of 1.5$^o$ between neighbour analyzers. The vertical arm houses only
one analyzer. Energy scans are done by varying the temperature of the
monochromator with respect to that of the analyzer crystals. The absolute
energy calibration between succesive scans is better than 1 meV. Each scan
took about 180~min, and each $(Q$-point spectrum has been obtained from the
average of 2 to 8 scans depending on h and the $Q$-transfer. The data have
been normalized to the intensity of the incident beam. The liquid lithium
uncapped container is made out of austenitic stainless steel with a
resistence heater, used to kept the liquid at $T=475$ K. The 20 mm long
sample, kept together by surface tension, was maintained in a 10$^{-6}$ bar
vacuum. The lithium has been loaded in an argon glove box. In the $Q-E$
region of interest, empty vacuum chamber measurements gave either the flat
electronic detector background of 0.6~counts/min or, at $9<Q<13$ nm$^{-1}$, 
a small elastic line due to scattering from the chamber kapton windows (each 
50 mm thick) which was subtracted from the data.

The IXS spectra of liquid lithium are reported in Fig.~1 at the indicated $Q$
transfer values. The low $Q$ data show the Brillouin triplet structure with
the energy of the inelastic peaks increasing with $Q$ up to a $Q$ value of
12 $nm^{-1}$. This value corresponds to $Q_M/2$, as deduced from the $S(Q)$
reported in Fig.~2. One can interpret, therefore, the dispersion up to $%
Q_M/2 $ as that of the longitudinal acoustic branch in the pseudo-first BZ.
Furthermore, similarly to what it is found in the second BZ of a crystal, we
observe that, also in liquid lithium, the energy of the acoustic modes
decreases with increasing $Q$ from $Q_M/2$ to $Q_M$. Increasing $Q$ above
$Q_M$, i.~e. in the "third" or higher BZs, one finds that the spectrum gets
incresingly broader and distinct peaks are no longer observable. At the
highest $Q$-values one finds that the $S(Q,E)$ becomes a symmetric peak
centered at energies larger than $E=0$. Beside the observation of dispersion
in a first and second pseudo-BZs, it is also important to note that the
broadening of the excitations monotonically increase with $Q$ to the extent
that towards the end of the second pseudo-BZ the inelastic features are no
longer showing a well defined peak.

The previous qualitative description can be made substantially more
quantitative considering that the dynamic structure factor $S(Q,E)$ can be
derived from the measured intensity, $I(Q,E)$, using the zeroth and first
moment sum rules for $S(Q,E)$: 
\begin{eqnarray}
m_{(0)}^S &=&\int S(Q,E)dE=S(Q), \\
m_{(1)}^S &=&\int ES(Q,E)dE=\hbar ^2Q^2/2M.
\end{eqnarray}
Considering that $I(Q,E)=\alpha (Q)\int dE^{\prime }S(Q,E^{\prime
})R(E-E^{\prime })$, where R(E) is the experimental resolution function
and $\alpha (Q)$ is a factor taking into account the scattering
geometries -
efficiencies and the lithium atomic form factor, the first moments of the
experimental data, $m_{(0)}^I$ and $m_{(1)}^I$, and those of the resolution
function, $m_{(0)}^R$ and $m_{(1)}^R$, are related to $m_{(0)}^S$ and
$m_{(1)}^S$ by: 
\begin{eqnarray}
m_{(0)}^I &=&\alpha (Q)m_{(0)}^Sm_{(0)}^R, \\
m_{(1)}^I &=&\alpha (Q)(m_{(0)}^Sm_{(1)}^R+m_{(1)}^Sm_{(0)}^R).
\end{eqnarray}
From the previous equation one derives that 
\begin{equation}
S(Q)=\frac{\hbar ^2Q^2}{2M}(m_{(1)}^I/m_{(0)}^I-m_{(1)}^R/m_{(0)}^R)^{-1}.
\end{equation}
This procedure has been utilised to put the $S(Q,E)$ on an absolute scale
using the experimentally determined $I(Q,E)$ and $R(E)$. The reliablitiy of
this procedure is shown in Fig.~2 where we obtain an excellent agreement
between the $S(Q)$ values obtained by Eq.~4 and those derived by Molecular
Dynamics (MD) simulation \cite{tullioMD}.

The possibility to express $S(Q,E)$ in absolute units allows to compare the
IXS data with the Gaussian lineshape expected for the $S(Q,E)$ when the
single particle limit is reached. This comparison is reported in Fig.~1b,
where each solid line represents the sum of two Gaussians: 
\begin{equation}
G(Q,E)=\frac{1}{\sqrt{2\pi}} \sum_{i=6,7} \frac{C_i}{\sigma_i} \exp{%
-(E-E_i)^2/2\sigma_i^2} 
\label{isotopi}
\end{equation}
where $C_6=0.08$ and $C_7=0.92$ are the natural abundances of the $^6$Li and 
$^7$Li isotopes, $E_i=\hbar^2 Q^2 /2 M_i$ their recoil energies and
$\sigma^2_i=\hbar^2 K_B T Q^2/ M_i$. The progressively better agreement
between
the data and $G(Q,E)$ with increasing $Q$ testifies the evolution towards
the single particle behaviour, which seems to be reasonably well reached at
the highest investigated $Q$ values and at the considered temperature.

The dispersion relation of the energy $\Omega (Q)$ of the inelastic signal
observed in Fig.~1 is obtained by determining the maximum of the current
spectra. This allows us to estimate $\Omega (Q)$ independently from any
specific model for the $S(Q,E)$. Examples of current spectra, obtained from
the spectra in Fig.1, are shown in Fig.~3. The values of $\Omega (Q)$ have
been determined by a parabolic fit of the maximum region in the Stokes side.
The values of $\Omega (Q)$ are reported in Fig.~4. In Fig.4a are also
reported the linear dispersion of the adiabatic sound velocity (dotted line)
and the parabolic dispersion expected for the single particle dynamics. We
observe that $\Omega (Q)$ is close to the adiabatic sound mode at the lowest 
$Q$ values, and it shows a positive dispersion before reaching the maximum
of the first pseudo-BZ. This is emphasized in Fig.~4b, where the low $Q$
region is expandend, and the inset reports directly the sound velocity
$v(Q)=\Omega (Q)/Q$. This behaviour confirms previous MD
\cite{rubidioMD,cesioMD,liMD} and experimental \cite{sinn}
data in a similar $Q$
region on lithium and other alkali liquid metals. With increasing $Q$
values, the points in Fig.~4a show not only a second pseudo-BZ, as already
pointed out in Fig.~1, but also a series of oscillations that dump out with
increasing $Q$ - here, $\Omega (Q)$ approaches the single particle line.
These oscillations are in anti-phase with the oscillations found in the $S(Q)
$ (see Fig.~2) and can, therefore, be associated to the local order in the
liquid. It is of great interest to be able to follow all the way from the
sound mode to the free particle regime the evolution of $\Omega (Q)$.
Beside the importantce of a unified picture in such a wide $Q$ region these
data provide the workbench for a quantitative theoretical analysis of the
shape of the $S(Q,E)$.

In conclusion, using IXS, we have been able to measure the collective
dynamics in a simple liquid from the collective modes regime, dominated by
sound-like excitations, all the way towards the impulse approximation
regime, dominated by single particle dynamics. This textbook result provides
a picture on how in a simple liquid the structural correlations induce
important deviation from the "continuum" picture used in the hydrodynamic
limit, and gives effects qualitatively similar to those as the BZs in
crystals.

We acknowledge valuable help of H.~Mueller from the Chemestry Laboratory at 
ESRF for his technical assistance during the sample manipulation.

\begin{center}
{\bf FIGURE CAPTIONS}
\end{center}

{\footnotesize { 

\begin{description}
\item  {FIG. 1 - Dynamic Structure Factor (Density Correlation Function) 
of Lithium at $T=475K$ measured by IXS and normalized according to the 
procedure discussed in the text. The spectra on the left column are taken
with a resolution $\Delta E =3$ meV using Si (9 9 9) reflection; Those
on the rigth column with $\Delta E =8.5$ meV using the Si (7 7 7) reflection.
In the latter column the comparison with a gaussian lineshape (full line) 
expected in the free particle limit is also reported.

\item  {FIG. 2 - Test of normalization accuracy: the $S(Q)$
values as extracted from IXS normalization (open symbols) are reported
toghether with computer simulation data (full line) \cite{tullioMD}.
The inset shows an enlargement of the small $Q$ region and the dotted
line is the $Q\rightarrow  0$ limit expected from the isothermal sound
velocity.

\item  FIG. 3 - Current Correlation Function ($J(Q,E)$) obtained from the 
$S(Q,E)$ spectra reported in Fig.1 as $J(Q,E)=E^2/Q^2 S(Q,E)$.

\item  FIG. 4 - Sound dispersion of Lithium. Fig 4a - The energy position of 
the Stokes Peak of the Current Correlation Function (open cirlce) are
reported toghether with the dispersin expected in two limiting cases:
low Q linear dispersion with hidrodynamic velocity ((dotted line) and
high Q (full line) parabolic dispersion of the impulse approximation.
These two dispersion curves have been compute assuming the presence of
both $^6$Li and $^7$Li in their
natural abundances (see text and Eq.~\ref{isotopi}).

Fig 4b - Detail of the `small Q` 
(below the First Sharp Diffraction Peak) region. The deviation of the
acoustic sound velocity from the adiabatic value (dotted line) at
increasing wavevectors
is observed (positive dispersion). 
}}
\end{description}

\end{document}